\documentclass[preprint,showpacs,preprintnumbers,amsmath,amssymb]{revtex4}

\usepackage{graphicx}% Include figure files
\usepackage{dcolumn}% Align table columns on decimal point
\usepackage{bm}% bold math

\begin{document}

\preprint{Submitted, 2008}

\title{Modeling Spacing Distribution of Queuing Vehicles in Front of a Signalized Junction Using Random-Matrix Theory}

\author{Xuexiang Jin, Yuelong Su, Yi Zhang, Li Li}
 \altaffiliation{Department of Automation, Tsinghua University, Beijing 100084, P. R. China}
 \email{li-li@mail.tsinghua.edu.cn}

\date{\today}% It is always \today, today,

\begin{abstract}
Modeling of headway/spacing between two consecutive vehicles has
many applications in traffic flow theory and transport practice.
Most known approaches only study the vehicles running on freeways.
In this paper, we propose a model to explain the spacing
distribution of queuing vehicles in front of a signalized junction
based on random-matrix theory. We show that the recently measured
spacing distribution data well fit the spacing distribution of a
Gaussian symplectic ensemble (GSE). These results are also compared
with the spacing distribution observed for car parking problem. Why
vehicle-stationary-queuing and vehicle-parking have different
spacing distributions (GSE vs GUE) seems to lie in the difference of
driving patterns.
\end{abstract}

\pacs{
      {05.20.Gg},%{Classical ensemble theory}
      {05.30.Ch},%{Quantum ensemble theory}
      {02.50.r},%{Probability theory, stochastic processes, and statistics}
    % end of PACS codes
} %end of abstract

\maketitle

\section{Introduction}
\label{intro}

The spacing is usually defined as the distance between two
successive vehicles measured from the same common feature of the
vehicles (e.g. rear axle, front bumper). Because the distribution of
spacing reflects the unmeasurable interaction forces or potentials
between vehicles that governs their motions, increasing
investigation are put into this field to reveal the complex dynamics
of vehicle traffic flow and explain some important phenomena, i.e.
phrase transition \cite{ChowdhurySantenSchadschneider2000},
\cite{Helbing2001}, \cite{MahnkeKaupuzsLubashevsky2005},
\cite{SchonhofHelbing2007}.

One interesting topic is to discuss spacing distributions observed
during the formation of and transitions between different vehicle
queues: static queues (vehicles parked on a line in the parking lot,
or vehicle queues fully-stopped in front of signalized
intersections), moving queues which may contain diversified
inter-arrival and inter-departure queuing interactions
\cite{SchonhofHelbing2007}, \cite{HelbingTreiberKesting2006},
\cite{ZhangWangWeiChen2007}. In this short paper, we will focus on
the not so popular static queues.

The vehicle parking problem was first introduced by Renyi in
\cite{Renyi1963} as: \emph{how many randomly parking motorists can
be accommodated on a line street of a given length on average}. The
most famous solution to this question is based on Random Sequential
Adsorption (RSA), which looks it as an irreversible process in which
particles are adsorbed sequentially and without overlap onto
randomly chosen positions on a surface. This 1D RSA problem can be
solved analytically when all the vehicles with the same equal
length, see \cite{TalbotTarjusVanTasselViot2000}, \cite{Lee2004},
\cite{RawalRodgers2005}. However, it is hard to directly apply this
method to other vehicle queues by definition.

Differently in \cite{Abul-Magd2006}, the random-matrix theory is
used to study the car-parking problem, where the nature of
interaction between the particles in a Dyson's Coulomb gas model is
assumed to be consistent with the tendency of the drivers to park
their cars near to each other and in the same time keep a distance
sufficient for departure. It was shown that the recently measured
gap-size distribution of parked cars in a number of roads in central
London can be well represented by the spacing distribution of a
Gaussian unitary ensemble.

For a similar purpose, in this paper, the Dyson's Coulomb gas model
is adopted to explain the formation of vehicle queues fully-stopped
in front of signalized intersections. We found a good agreement
between the empirical data and the spacing distribution for Gaussian
symplectic ensemble of random matrices.

\section{Coulomb Gas Model and Queuing Vehicles}
\label{sec:1}

\begin{figure}[h]
\centerline{\includegraphics[height=2.3in]{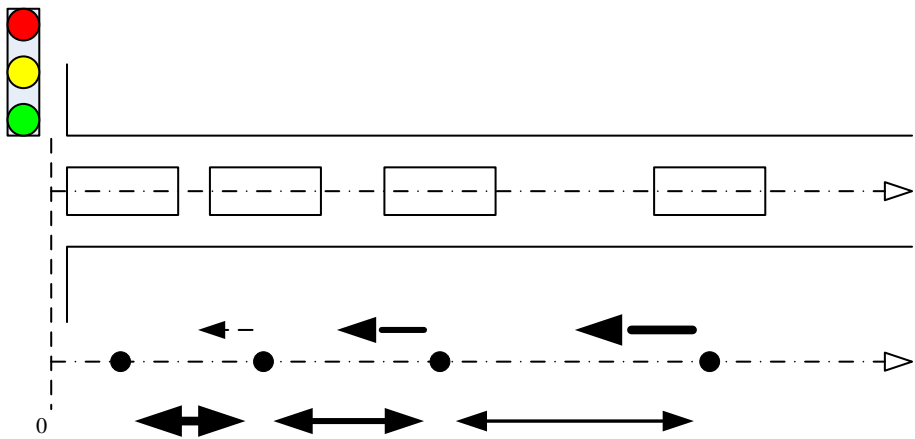}}
\caption{Queuing vehicles in front of a signalized junction analogue
to the Dyson Gas.} \label{fig1}
\end{figure}

Considers a gas of $N$ charges whose positions are denoted by $x_1$,
$x_2$, ..., $x_N$ and these charges are free to move on the line $0
< x < + \infty$, see Fig.~\ref{fig1}. Suppose the potential energy
of this Coulomb gas is given as \cite{Dyson1962}

\begin{equation}
\label{equ:1} V = \frac{1}{2} \sum_i x_i^2 - \sum_{i<j} \ln | x_i -
x_j |
\end{equation}

where the first term in (\ref{equ:1}) represents a harmonic
potential that attracts each charge independently towards the
coordinate origin; and the second term represents an electrostatic
repulsion between each pair of these charges.

Let $P(s)$ denote the nearest-neighbor spacing distribution of these
charges. The accurate solution of $P(s)$ is not easy to find.
However, for such systems, the probability density function for the
position of the charges can be approximately calculated by using the
so called Wigner surmise \cite{GuhrMuller-GroelingWeidenmuller1998},
\cite{Mehta2004}, \cite{DiengTracy0603543v1}.

Suppose the gas is in thermodynamical equilibrium at temperature
$T=1/(K \beta)$, where $K$ is the Boltzmann constant. The
probability density function can be given as below by the Boltzmann
factor obtained from the Gibbs-Boltzmann canonical distribution by
integration over the momenta of the particles.

\begin{equation}
\label{equ:2} P(x_1, x_2, ..., x_N) = C e^{-\beta V}
\end{equation}

Combining (\ref{equ:1}) and (\ref{equ:2}), the Wigner surmise
solutions for $P(s)$ can be gotten by taking the additional
assumption of $\beta$. The role of this inverse temperature $\beta$
denotes the level-repulsion power of the matrices eigenvalues.
Particularly, they are suggested by Wigner as below respectively.
For $\beta = 0$, we get the well-known Poisson Ensembles (PE)

\begin{equation}
\label{equ:3} P_{PE}(s) = e^{-s}
\end{equation}

for $\beta = 1$, we get Gaussian Orthogonal Ensembles (GOE)

\begin{equation}
\label{equ:4} P_{GOE}(s) = \frac{1}{2} \pi s e^{- \frac{\pi}{4} s^2}
\end{equation}

for $\beta = 2$, we get Gaussian Unitary Ensemble (GUE)

\begin{equation}
\label{equ:5} P_{GUE}(s) = \frac{32}{\pi^2} s^2 e^{- \frac{4}{\pi}
s^2}
\end{equation}

, and for $\beta = 4$, we get Gaussian Orthogonal Ensembles (GSE)

\begin{equation}
\label{equ:6} P_{GSE}(s) = \frac{2^{18}}{3^6 \pi^3} s^4 e^{-
\frac{64}{9 \pi} s^2}
\end{equation}

Similar as discussed in \cite{Abul-Magd2006}, we can abstract the
movements of vehicles of different size into point particles,
because we are only interested in the spacing distribution here. A
natural guess for the queuing dynamics of vehicles in front of a
signalized junction is that the system can also be approximately
formulated into this Coulomb model.

The single-particle term in (\ref{equ:1}) can be viewed to reflect
the tendency of driving closer and the repulsive two-body term in
(\ref{equ:1}) indicate the tendency of maintaining the safe
distance. Analogously, the basic instinct of a driver is to maintain
a small and safe gap between him/her and his/her leading vehicle,
especially when he/she is queuing. The superposition of these two
potentials, which appears an overall repulsion for small spacings
and attraction for the large ones, expresses the fact that it is
unlikely to see too small or too large spacings between queuing
vehicles. However, no one can always keep an ideal headway due to
disturbances (unexpected acceleration/deceleration of the leading
vehicle, occasional absence of mind, etc.). Thus, these vehicles
(particles) are perturbed by environment simultaneously.

Noticing the above analogue and inspiriting by the report that the
spacing distribution of vehicle parking is in agreement with GUE
type distribution, we conjecture that the empirical spacing
distribution of a queuing vehicle systems might fit one kind of
Wigner surmises (\ref{equ:3})-(\ref{equ:6}), too.

\section{Comparison with Empirical Results}
\label{sec:2}

To test this conjecture, we collected 700 sample spacings of queuing
vehicles in front of several different signalized junctions in
Beijing, China. Some details about data collection can be found in
\cite{SuWeiChengYaoZhangLiZhangLi2008}.

\begin{figure}[h]
\centerline{\includegraphics[height=4.3in]{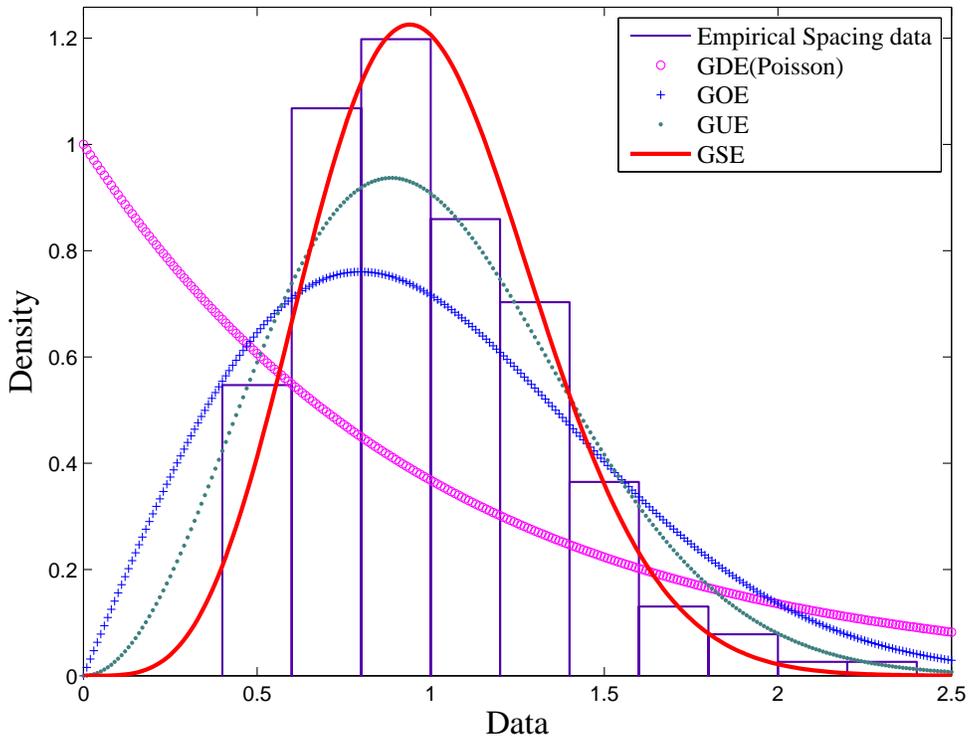}} \caption{The
\textbf{normalized} spacing distribution of queuing vehicles in
front of several signalized junctions in Beijing compared with the
theoretical spacing distribution for Poisson, GOE, GUE, GSE.}
\label{fig2}
\end{figure}

The average spacing size observed is 1.43m here. Fig.~\ref{fig2}
shows the probability distribution function $P(s)$ in a form of
\textbf{normalized} histogram, where the values of the $x$-axis is
defined as the ratio of the spacing to the mean value. This
modification is introduced to make the data comparable with the
nearest-neighbor spacing distribution for system (\ref{equ:1}).

In the Coulomb gas model, the inverse temperature $\beta$ of the gas
characterizes the degree of repulsion. Fig.~\ref{fig2} shows the
theoretical spacing distribution curves for $P_{PE}(s)$,
$P_{GOE}(s)$, $P_{GUE}(s)$ and $P_{GSE}(s)$ as well as the empirical
spacing distribution histogram. We can see that different from the
vehicle-parking scenarios, the vehicle-queuing scenarios well fit
the GSE type model instead of GUE type model, although in these two
scenarios, drivers all aim to driving close enough but not too
close. This suggests that the vehicles queuing process at a
signalized junction can be added to the long list of the systems
with RMT-like fluctuations.

An interesting question is why the spacing distributions of
vehicle-parking and vehicle-queuing are different. A reasonable
explanation is this difference comes from the dissimilar driving
patterns. In vehicle-parking scenarios, drivers would like to try
several times and move back-and-forth to adjust the gaps so as to
park to an ``ideal'' position; while in vehicle-queuing scenarios,
driver won't be able to drive back. This phenomenon can be
interpreted as: in the vehicle-parking scenarios, the repelling
force from the neighboring vehicles is relatively ``loose'';
however, in the vehicle-queuing scenarios, the repel force from the
neighboring vehicles is a kind of ``rigid''. As pointed out in [7],
such a difference will result in the different values of inverse
temperature $\beta$. At low temperatures ( $\beta$ is bigger), the
charges tend to be regularly spaced in a crystalline lattice
arrangement and the randomness of the positions of the charges are
small. At higher temperatures ( $\beta$ is smaller), the
fluctuations of the charges become tenser. Thus, for vehicle-parking
scenarios, we get $\beta = 2$ and for vehicle-queuing, we get $\beta
= 4$.

We also guess that in different cities, the spacing of vehicle
queues may still hold GSE type distribution but may have different
mean value. Further experiments will be carried out to test this
guess soon. Any vehicles queuing data collected in cities other than
Beijing are welcome and highly appreciated.

\begin{acknowledgments}

This work was supported in part by National Basic Research Program
of China (973 Project) 2006CB705506, Hi-Tech Research and
Development Program of China (863 Project) 2006AA11Z229, and
National Natural Science Foundation of China 60374059, 50708055.

\end{acknowledgments}

%
% BibTeX users please use
% \bibliographystyle{}
% \bibliography{}

\begin{thebibliography}{}

\bibitem{ChowdhurySantenSchadschneider2000}
D. Chowdhury, L. Santen, A. Schadschneider, \textit{Phys. Rep.}
\textbf{329}, 199 (2000).
% Statistical physics of vehicular traffic and some related systems

\bibitem{Helbing2001}
D. Helbing, \textit{Rev. Mod. Phys.} \textbf{73}, 1067 (2001).
% Traffic and related self-driven many-particle systems

\bibitem{MahnkeKaupuzsLubashevsky2005}
R. Mahnke, J. Kaupuzs, I. Lubashevsky, \textit{Phys. Rep.}
\textbf{408}, 1 (2005).
% Probabilistic description of traffic flow

\bibitem{SchonhofHelbing2007}
M. Schonhof, D. Helbing, \textit{Transp. Sci.} \textbf{41}, 135
(2007).
% Empirical features of congested traffic states and their implications for traffic modeling

\bibitem{ZhangWangWeiChen2007}
G. Zhang, Y. Wang, H. Wei, Y. Chen, in \textit{Transportation
Research Board Annual Meeting CD}, 2007.
% Examining headway distribution models using urban freeway Loop Event Data

\bibitem{HelbingTreiberKesting2006}
D. Helbing, M. Treiber, A. Kesting, \textit{Physica A} \textbf{36},
62 (2006).
% Understanding interarrival and interdeparture time statistics from interactions in queuing systems

\bibitem{Renyi1963}
A. Renyi, \textit{Sele. Trans. Math. Stat. Probab.} \textbf{4}, 203
(1963).
% On a one-dimensional problem concerning random space filling

\bibitem{TalbotTarjusVanTasselViot2000}
J. Talbot, G. Tarjus, P. R. Van Tassel, P. Viot, \textit{Colloids
and Surf. A} \textbf{165}, 287 (2000).
% From car parking to protein adsorption: An overview of sequential adsorption processes

\bibitem{Lee2004}
J. W. Lee, \textit{Physica A} \textbf{331}, 531 (2004).
% Reversible random sequential adsorption on a one-dimensional lattice

\bibitem{RawalRodgers2005}
S. Rawal, G. J. Rodgers, \textit{Physica A} \textbf{346}, 621
(2005).
% Modelling the gap size distribution of parked cars

\bibitem{Abul-Magd2006}
A.Y. Abul-Magd, \textit{Physica A} \textbf{368}, 536 (2006).
% Modelling gap-size distribution of parked cars using random-matrix theory

\bibitem{Dyson1962}
F. J. Dyson, \textit{J. Math. Phys.} \textbf{3}, 140 (1962a).
% Statistical theory of the energy levels of complex systems I

\bibitem{GuhrMuller-GroelingWeidenmuller1998}
T. Guhr, A. Muller-Groeling, H. A. Weidenmuller, \textit{Phys. Rep.}
\textbf{299}, 189 (1998).
% Random-matrix theories in quantum physics: Common concepts

\bibitem{Mehta2004}
M. L. Mehta, \textit{Random Matrices}, 3rd edition, (Academic Press,
Boston, 2004).

\bibitem{DiengTracy0603543v1}
M. Dieng, C. A. Tracy. arXiv:math/0603543v1
% Application of random matrix theory to multivariate statistics

\bibitem{SuWeiChengYaoZhangLiZhangLi2008}
Y. Su, Z. Wei, S. Cheng, D. Yao, Y. Zhang, L. Li, Z. Zhang, Z. Li,
in \textit{Transportation Research Board Annual Meeting CD}, (2008).
% Departure headways of mixed traffic flow at signalized intersections: Distributions, simulations and validations

% etc
\end{thebibliography}

\end{document}